\documentclass[12pt]{article}

\usepackage[utf8]{inputenc}
\usepackage{tabularx,booktabs,colortbl}
\usepackage[frozencache, cachedir=.]{minted}
\renewcommand\listingscaption{Figure}
\usepackage[dvipsnames]{xcolor}
\usepackage{color}
\usepackage{soul}
\definecolor{LightGray}{gray}{0.9}
\usepackage{tikz}
\usetikzlibrary{tikzmark}

\usepackage{amsmath,array,graphicx}
\usepackage{textcomp}
\usepackage{gensymb}
\usepackage{float}
\usepackage{graphicx}
\usepackage{textgreek}
\usepackage{multicol, multirow}
\usepackage[many]{tcolorbox}
\usepackage{makecell}

\usepackage[labelfont=bf]{caption}
\usepackage[textfont=it]{caption}
\usepackage[margin=0.75in]{geometry}
\usepackage[superscript]{cite}
\usepackage{tabularx}
\definecolor{box_green}{RGB}{0,188,0}
\usepackage{subcaption}
\newtcolorbox{promptbox}{
  unbreakable,
  enhanced,
  colback=white,
  colframe=box_green,
  arc=0pt,
  outer arc=0pt,
  title=Generating a Methods Section,
  fonttitle=\bfseries\sffamily\large,
  colbacktitle=box_green,
  attach boxed title to top left={},
  boxed title style={
    enhanced,
    skin=enhancedfirst jigsaw,
    arc=3pt,
    bottom=0pt,
    interior style={fill=box_green}
  }
}

\newtcolorbox{promptquotebox}[1]{
  unbreakable,
  enhanced,
  colback=white,
  colframe=box_green,
  arc=0pt,
  outer arc=0pt,
  title=#1,
  fonttitle=\bfseries\sffamily\large,
  colbacktitle=box_green,
  attach boxed title to top left={},
  boxed title style={
    enhanced,
    skin=enhancedfirst jigsaw,
    arc=3pt,
    bottom=0pt,
    interior style={fill=box_green}
  }
}

\newtcolorbox{GPT-4}{
  title=GPT-4,
  breakable,
  enhanced,
  colback=lightgray,
  colframe=black,
  arc=0pt,
  outer arc=0pt,
  fonttitle=\bfseries\sffamily\large,
  colbacktitle=white,
  attach boxed title to top left={},
  boxed title style={
    enhanced,
    skin=enhancedfirst jigsaw,
    arc=3pt,
    bottom=0pt,
    interior style={fill=black}
  }
}

\DeclareCaptionFormat{suboverlay}{\gdef\subcapoverlay{(\thesubfigure) #3\par}}
\DeclareCaptionStyle{suboverlay}{format=suboverlay}

\newcommand{\subcaptionOverlay}[1]{%
  \subcaption{}%
  \begin{tikzpicture}
    \node [inner sep=0,anchor=north west]at (-2ex,-2ex) (image) {#1};
    \draw node [black] {\subcapoverlay};
  \end{tikzpicture}%
}

\title{GPT-4 as an interface between researchers and computational software: improving usability and reproducibility}

\author{Juan C. Verduzco$^{a,*}$, Ethan W. Holbrook$^{a,*}$, and Alejandro Strachan$^a$}

\date{
$^a$ School of Materials Engineering and Birck Nanotechnology Center \\
Purdue University, West Lafayette, Indiana, 47907 USA
\\
* Authors contributed equally to this work.
}

\begin{document}

\maketitle

\begin{abstract}

Large language models (LLMs) are playing an increasingly important role in science and engineering. For example, their ability to parse and understand human and computer languages makes them powerful interpreters and their use in applications like code generation are well-documented. We explore the ability of the GPT-4 LLM to ameliorate two major challenges in computational materials science: i) the high barriers for adoption of scientific software associated with the use of custom input languages, and ii) the poor reproducibility of published results due to insufficient details in the description of simulation methods. We focus on a widely used software for molecular dynamics simulations, the Large-scale Atomic/Molecular Massively Parallel Simulator (LAMMPS), and quantify the usefulness of input files generated by GPT-4 from task descriptions in English and its ability to generate detailed descriptions of computational tasks from input files. We find that GPT-4 can generate correct and ready-to-use input files for relatively simple tasks and useful starting points for more complex, multi-step simulations. In addition, GPT-4's description of computational tasks from input files can be tuned from a detailed set of step-by-step instructions to a summary description appropriate for publications. Our results show that GPT-4 can reduce the number of routine tasks performed by researchers, accelerate the training of new users, and enhance reproducibility.
\end{abstract}


\section*{Introduction}
\label{introduction_section}

Artificial intelligence (AI) is playing an increasingly important role in science and engineering. Applications range from design of experiments \cite{stein2019progress} and self-driving labs \cite{hase2019next}, to enhancing physics-based simulations \cite{bartok2010gaussian} and the discovery of physical laws \cite{desai2021parsimonious}. Supervised learning models have been successful in accelerating drug \cite{nigam2021beyond} and materials discovery \cite{xue2016accelerated} while the integration of physical laws into neural networks has proven useful to discover and solve partial differential equations\cite{raissi2019physics}. Another exciting application of AI tools builds on progress in natural language processing (NLP) tasks to extract information from scientific publications and reports, either to summarize results, answer questions, make inferences, or extract data of interest \cite{qin2023chatgpt, paraskevopoulos2022using, polak2023extracting, polak2023flexible}. A recent breakthrough in NLP was the introduction of transformer architectures \cite{vaswani2017attention}, which process entire input sequences at once, rather than sequentially. This advancement facilitates capturing dependencies between words in a sentence and resulted in the development of large language models (LLMs). These models, underlying many popular chatbots, have shown a remarkable ability to solve a wide range of problems \cite{openAI2023GPT4} and have been credited with ``sparks of general intelligence'' \cite{bubeck2023sparks}.  This paper explores the potential of the GPT-4 LLM to serve as an interface between a researcher and scientific software, with the goal of streamlining and accelerating routine tasks, training novice users, and improving reproducibility. 

Among other models, the Generative Pre-trained Transformer (GPT)\cite{radford2018improving} model from OpenAI and the Bidirectional Encoder Representations from Transformers (BERT) model from Google \cite{devlin2018bert} have revolutionized the field of AI language modeling. There are ongoing efforts to explore the use of these models across various domains and results indicate a significant potential for diverse applications such as virtual assistants, automated content creation, document analysis, and machine translation \cite{openAI2023GPT4}.
GPT-4 is a multimodal model capable of accepting both images and text \cite{openAI2023GPT4} as inputs. However, the full model has not been widely distributed at the time of writing; an early text-only version is available through OpenAI's ChatGPT and via a paid API \cite{bubeck2023sparks}. Detailed information regarding GPT-4's construction is currently scarce but general details and numerous performance metrics are available \cite{openAI2023GPT4, bubeck2023sparks}. The model is trained on an extensive collection of web text data, with the primary objective of predicting the next word (or token) in a sentence or sequence of words \cite{openAI2023GPT4}. This allows the model to learn patterns, allowing it to generate coherent and contextually appropriate text based on its input. As discussed above, this approach has led to the  remarkable capacity of LLMs to perform complex tasks in many different fields.

Unsurprisingly, LLMs are also being evaluated for applications in materials science. Morgan and collaborators used LLMs to extract properties and associated metadata from scientific publications \cite{polak2023flexible}. They developed an approach that combines text processing, LLMs, and minimal human intervention suitable for mid-sized data extraction problems. LLMs have also been used to generate the building blocks required to run density functional theory (DFT) calculations from text descriptions\cite{Hong2023}. This research primarily focused on the generation of atomic structure files for a specific problem and iteratively improving on them. The creation of input files for standard DFT simulations was also addressed, testing the ability of ChatGPT to produce a mostly functional script resembling examples found in the code's documentation \cite{Hong2023}. However, this publication has a narrow focus and does not quantify the accuracy of the resulting scripts. 

This paper focuses on GPT-4's ability to interface with LAMMPS \cite{plimpton1995fast}, a widely used code for molecular dynamics (MD). We evaluate input files generated by GPT-4 from three task descriptions created by the authors and two obtained from the literature and its ability to generate human readable descriptions of a computational task from a LAMMPS input file. We find that GPT-4 creates working and accurate input files for relatively simple and well described tasks. This achievement by itself is quite remarkable; GPT-4 selects the correct crystal structure for the material of interest and the correct thermostat to create the desired thermodynamic ensemble, among other simulation details. For complex tasks like simulating spall failure following a high-velocity impact simulation, GPT-4 creates input files that, while requiring modification, provide an excellent starting point and significantly reduce the time required to create a working input file.  Finally, given a LAMMPS input file, GPT-4 is capable of generating a detailed description of the simulation performed and, if prompted, can provide a summary statement that is appropriate to be included in the methods section of a publication.

\section*{Methods}
\label{methods_section}

We investigate GPT-4's ability to generate LAMMPS input files from the description of MD tasks in English, as well as the inverse problem of explaining what a LAMMPS input file does. Our goals are to reduce the number of routine tasks performed by researchers, accelerate the training of new users, and enhance reproducibility. To address the first challenge, we designed two tasks. Task 1 assesses GPT-4 generated LAMMPS input files from three task descriptions of increasing complexity created by the authors. Task 2 tests the creation of LAMMPS input files based on the methods section of two publications. To address the second challenge, Task 3 prompts GPT-4 to generate comprehensive and summary human-readable descriptions of computational tasks provided with a LAMMPS input file obtained from literature. These tasks evaluate GPT-4's understanding of the LAMMPS input language but also its ability to generalize, which enables it to perform tasks combining information about various topics, to understand complex procedures, and their translation into instructions \cite{bubeck2023sparks}. 

For each prompt, we asked GPT-4 to generate ten LAMMPS input files and we executed them using LAMMPS version 29 Oct 2020 for Task 1 and Task 2.2 and version 15 Sep 2022 for Task 2.1 which required a separate set of packages. The generated input files are evaluated according to the metrics outlined in Table \ref{metrics_table}. We define ``script validity'' as the fraction of input files that execute in their entirety without any modification and ``script accuracy'' refers to the subset of those files that successfully performed the task requested. ``Consistency" measures the average pairwise similarity between the files generated and ``usefulness" denotes the average similarity between the GPT-4 generated files and a {\it gold standard} input file created by the authors. We measure similarity between input files using the cosine similarity criteria. To account for the non-numerical and non-key-pair values, each file is transformed into term-frequency vectors \cite{HAN201239}. Thus, the cosine similarity metric identifies common keywords. We note that quantifying the similarity between LAMMPS input files is challenging due to the custom nature of the LAMMPS input language. For example the order of commands may or may not affect the simulation. In addition, similar key-value pairs can result in drastically different physics simulations (i.e. \textit{fix 1 all npt/sphere temp 300.0 300.0 100.0} vs \textit{fix 1 all nvt temp 300.0 300.0 100.0}).

\begin{table*}[htbp]
    \caption{Definition of performance metrics to evaluate GPT-4 LAMMPS input files.}
    \label{metrics_table}
    \begin{center}
        \begin{tabular}{cccc}
            \toprule
            \textbf{\thead{Type}} & \textbf{\thead{Metric}} & \textbf{Equation} & \textbf{{Description}} \\
            \hline
            \multirow{3.2}{*}{Accuracy}& Script Validity & 
                \parbox{3cm}{\begin{equation*}
                    \frac{Files_{RUN}}{10}
                \end{equation*}}
            & \makecell{What percent of the generated files \\ can be run in LAMMPS?}\\
            \cmidrule{2-4}
            & Simulation Accuracy & 
                \parbox{3cm}{\begin{equation*}
                    \frac{Files_{CORRECT}}{10}
                \end{equation*}}
            & \makecell{What percent of the generated files \\ accurately follow the prompt directions?}\\
            \midrule
            \multirow{3.2}{*}{Reliability} & Consistency & 
                \parbox{3cm}{
                \begin{equation*}
                \begin{split}
                    Sim = \frac{\vec{a} \cdot \vec{b}}{\left\Vert  \vec{a} \right\Vert \cdot \left\Vert  \vec{b} \right\Vert} \\
                    \vec{a}, \vec{b} \Rightarrow files
                \end{split}
                \end{equation*}}
            & \makecell{How different are files generated \\ for the same prompt?}\\
            \cmidrule{2-4} & Usefulness & 
                \parbox{3cm}{
                \begin{equation*}
                \begin{split}
                    Sim = \frac{\vec{a} \cdot \vec{G}}{\left\Vert  \vec{a} \right\Vert \cdot \left\Vert  \vec{G} \right\Vert} \\
                    \vec{a} \Rightarrow file \\ 
                    \vec{G} \Rightarrow Gold
                \end{split}
                \end{equation*}}&\makecell{How close to a correct solution \\ are the generated files?}\\           
            \bottomrule
            
        \end{tabular}
    \end{center}
\end{table*}

\section*{Results}
\label{tasks_section}

\subsection*{Task 1: LAMMPS input file generation from authors' descriptions}
\label{task1_section}

GPT-4 (version March 23, 2023) was tasked with the creation of LAMMPS input files from an English description of the desired simulation. We created three descriptions of MD tasks of increasing complexity, see box below, and used them as input to GPT-4 through ChatGPT with this prompt: Make a LAMMPS script based off this method description: `` \textit{Prompt} ''. Prompt engineering on distinct ways to request the creation of the file is out of the scope of this work. Ten input files generated were generated for each prompt and analyzed as discussed above after discarding the many useful comments added by GPT-4 to explain the workings of the LAMMPS commands. 

\begin{promptquotebox}{Prompt:Task 1}
\textit{We used molecular dynamics with LAMMPS to equilibrate an Al sample under isobaric, isothermal conditions (NPT ensemble) at 300 K and 1 atm for 500 ps. The initial conditions were obtained by replicating the fcc unit cell 5 times in each direction. We used Nose-Hoover thermostat and barostat with relaxation timescales of 0.1 and 1 ps, respectively. All MD simulations were performed using LAMMPS. Atomic interactions were obtained using an embedded atom model developed by Ercolessi and Adams [1] obtained from OpenKIM.org. [1] EAM potential (cubic natural spline tabulation) for Al developed by Ercolessi and Adams (1994) v002. OpenKIM. 2018. doi:10.25950/376e3e7e.}
\end{promptquotebox}

\begin{promptquotebox}{Prompt:Task 2}
\textit{We characterized the melting of a bulk Ni sample using molecular dynamics with LAMMPS. The initial condition was obtained by replicating the Ni unit cell 10 times in each direction. Initial velocities were drawn from the Maxwell-Boltzmann distribution at 600 K. The system was heated from 300 K to 2500 K continuously, at a rate of 10 K per ps under isothermal and isobaric conditions at 1 atm. Interactions were described using an embedded atom model developed by Mishin et al. in 1999 [1] obtained from OpenKIM.org. [1] EAM potential (LAMMPS cubic hermite tabulation) for Ni developed by Mishin et al. (1999) v005. OpenKIM; 2018. doi:10.25950/a88dfc37.}
\end{promptquotebox}

\begin{promptquotebox}{Prompt:Task 3}
\textit{We simulate spall failure on Nb single crystals using high-velocity impact simulations using molecular dynamics (MD) with the LAMMPS code. We simulate the impact of a projectile on a target with a relative velocity of 2 km/s. The projectile is obtained by replicating the Nb BCC unit cell 20 times along the [100], [010], and [001]; the target is longer along the shock direction and is obtained by replicating the BCC unit cell 20 times along [100] and [010] and 40 times along [001]. We apply periodic boundary conditions along the directions normal to the impact direction, [001], and free boundaries along [001]. A gap of 1.5 nm initially separates the target and projectile. The system is equilibrated at 300 K for 100 ps using isothermal, isochoric MD. An impact velocity of 2 km/s is added to the thermal velocities to all the atoms in the projectile along [001] in the direction of the target. Adiabatic MD is used to simulate the impact and subsequent expansion. All atomic interactions are described using an EAM potential developed by Fellinger et al. [1] and downloaded from openKIM [2]. [1] Fellinger MR, Park H, Wilkins JW. Force-matched embedded-atom method potential for niobium. Physical Review B. 2010Apr;81(14):144119. doi:10.1103/PhysRevB.81.144119
    [2] https://doi.org/10.25950/befb2eea.}
\end{promptquotebox}

We find the performance of GPT-4, summarized in Table \ref{task1_metrics}, remarkable. A vast majority of the input files generated for Prompts 1 and 2 (18/20) resulted in an error-free LAMMPS simulation without any modification and all but four of those produced the exact task requested. Figure  \ref{fig1:annotatedcode} shows a representative input script for Prompt 1; highlights indicate elements that are consistently correct (green), correct commands with values that vary from file to file (yellow) and an example of an incorrect statement (red). A similar breakdown of representative code for Prompt 2 is included in the supplementary information as Figure S\ref{figs1:annotatedcodeP2}. These input files consistently include non-trivial steps required to setup a meaningful MD simulation following LAMMPS's ad-hoc input language. These include setting up an atomistic model based on the material requested and an energy expression to desribe atomic interactions as well as selecting the appropriate barostat or thermostat to achieve the thermodynamic conditions in the prompt. The third input file is significantly more complex than the first two, it calls for the creation of two blocks of material and an impact simulation after thermalization. While none of the input files generated by GPT-4 work without modification, they provide an excellent starting point and significantly simply the process of creating the desired input file. The remainder of the section describes the resulting files in more detail, highlighting aspects that are generally captured correctly and common errors.

\begin{table}[ht]
\caption{Performance metrics for GPT-4 generated input files from different prompts describing a molecular dynamics simulation.}
\label{task1_metrics}
\vskip 0.3in
\begin{center}
\begin{tabular}{lccc|ccr}
\toprule
Metric & \tikzmark{a} Prompt 1 & Prompt 2 & Prompt 3 \tikzmark{c} & \makecell{Montes de Oca \\ Zapiain et al\cite{montes2022training}} & \makecell{Strachan \\ and Dorso \cite{strachan1998caloric}}  \\
\midrule
Script Validity          & 10/10 & 8/10 & 0/10 & 0/10 & 0/10\\
Simulation Accuracy      & 9/10  & 4/10 & 0/10 & 0/10 & 0/10\\
Consistency &  0.87  &  0.85 & 0.81  & 0.77 & 0.61\\
Usefulness &  0.88 & 0.76 & 0.81 & 0.74 & 0.67\\

\bottomrule
\end{tabular}

\begin{tikzpicture}[remember picture,overlay]
\draw[-latex] ([ shift={(0,5ex)}]pic cs:a) -- node[above] {Complexity} ([ shift={(0,5ex)}]pic cs:c);
\end{tikzpicture}

\end{center}
\vskip -0.1in
\end{table}

% CRYSTAL STRUCTURE, LATTICE PARAMETERS, BOUNDARY CONDITIONS

\begin{listing}[H]
\begin{minted}
[
frame=lines,
framesep=2mm,
baselinestretch=1.2,
bgcolor=LightGray,
escapeinside=||,
fontsize=\footnotesize,
linenos
]
{python}
|\colorbox{green}{units metal}|
|\colorbox{SpringGreen}{dimension 3}|
|\colorbox{green}{boundary p p p}|

|\colorbox{green}{atom\_style atomic}|
|\colorbox{green}{lattice fcc 4.05}|

|\colorbox{green}{region box 0 5 0 5 0 5} \colorbox{SpringGreen}{units lattice}|
|\colorbox{green}{create\_box 1 box}|
|\colorbox{green}{create\_atoms 1 box}|

|\colorbox{red}{replicate 5 5 5}|

|\colorbox{green}{pair\_style} \colorbox{yellow}{<eam/fs> or <eam/alloy>}|
|\colorbox{green}{pair\_coeff * *} \colorbox{yellow}{<potential file name>} \colorbox{green}{Al}|
|\colorbox{SpringGreen}{mass 1 26.9815}|

|\colorbox{SpringGreen}{neighbor 2.0 bin}|
|\colorbox{SpringGreen}{neigh\_modify delay 10}|

|\colorbox{green}{velocity all create 300} \colorbox{yellow}{<random seed>} \colorbox{SpringGreen}{mom yes} \colorbox{yellow}{rot yes} \colorbox{yellow}{dist <uniform> or <gaussian>}|

|\colorbox{green}{fix 1 all npt temp 300 300 0.1 iso 1 1 1}|
|\colorbox{green}{thermo} \colorbox{yellow}{<thermo steps>}|
|\colorbox{green}{thermo\_style} \colorbox{yellow}{<custom step temp pe etotal press vol>}|

|\colorbox{green}{timestep 0.001}|
|\colorbox{green}{run 500000}|
|\colorbox{SpringGreen}{unfix 1}|

|\colorbox{SpringGreen}{dump, print, write\_data, ...}|

\end{minted}
\caption{Annotated example of a LAMMPS input file from Prompt 1. \colorbox{green}{Always added correct commands}. \colorbox{SpringGreen}{Often added correct commands that match default values}. \colorbox{yellow}{Values that were generated with different options in some files}. \colorbox{red}{Incorrect commands}.  }
\label{fig1:annotatedcode}
\end{listing}

{\bf Model Selection.} All the files generated select the correct crystal structure (FCC for Al and Ni and BCC for Nb) and lattice parameters even when these values are not explicitly included in the prompt (except for Nb being BCC). It is unclear whether the model infers this information or it is selected from similar input files seen during training. The input files correctly set the boundary conditions as periodic for Prompts 1 and 2 and, remarkably, 9/10 input files for Prompt 3 correctly set up a free boundary along the Z-direction. 
% SETTING UP INTERATOMIC POTENTIALS

{\bf Interatomic potentials.} LAMMPS input files always refer to separate files for the interatomic potential required for the simulation and the file names vary between different files obtained from the same prompt. These names are placeholders, as GPT-4 consistently included warnings about the need to add the correct potential file with the correct name. For instance, in Prompt 1, the file name for the potential varied between different files, with names such as ''Al99.eam.alloy", ''Al.eam.alloy", and ''Al\_Ercolessi\_Adams.eam.alloy". The ''Al99.eam.alloy" filename coincides with the Al potential by Mishin and Farkas \cite{mishin1999interatomic} available on OpenKim \cite{tadmor2011potential}. However, none of the files matched the correct filename for the potential ''Al\_EA.eam.alloy" for the Ercolessi and Adams \cite{ercolessi1994interatomic} EAM potential. Interestingly, GPT-4 reference Finnis-Sinclair form of the Embedded Atom Model (EAM) is some cases and the standard EAM form in others; regardless of the form, the simulations are identical. 
% ENSEMBLE, SIMULATION LENGTH % CONSISTENCY, USEFULNESS

{\bf Ensemble and simulation time.} GPT-4 correctly interprets the prompts regarding the desired thermodynamic ensemble requested and uses the LAMMPS \textit{fix} command to setup the correct simulation in the majority of the cases. In all cases for Prompts 1 and 2, GPT-4 selects the correct ensemble and Nos\'{e}-Hoover thermostats and barostats. Prompt 3 requests a thermalization using isothermal-isochoric MD (NVT ensemble) followed by an NVE simulation of the process of impact. 8/10 of the input files use appropriate thermostats for the initial simulation (interestingly various approaches are used to accomplish this: Berendsen, Nos\'{e}-Hoover, and Langevin thermostats). The impact simulations call for an NVE simulation and GPT sets this up correctly in 9/10 cases. We note that setting up thermostats or barostats requires the non-trivial step of selecting damping parameters (relaxation times); values close to 0.1 ps are considered ideal for thermostatting under most conditions and a value around 1 ps is appropriate for barostats. GPT-4 uses appropriate relaxation times 23 out of the 27 times a Nos\'{e}-Hoover thermostat is used and 19/20 in the case of barostats. The length of the simulation, the product between time step and number of MD steps, is always accurate for Prompt 1 where it is explicitly declared, and is inaccurate in 3/10 files generated for Prompt 2 where it has to be inferred from the heating rate. For Prompt 3, a spall simulation, the description did not specify a total simulation time. This is common practice, and it is assumed that the simulation time would be long enough for spall failure to fully develop in the case of strong enough shocks. This time can be estimated from the speed of sound of the material, or more accurately, from knowledge of the shock speed as a function of particle velocity. GPT-4 appears to select a standard simulation time. 

{\bf Consistency and accuracy.} We now turn our attention to the consistency and accuracy metrics, see Table \ref{metrics_table}. The cosine similarity metric indicates that the input files for each prompt are closely related to each other, with similarity scores between 0.81 and 0.87. To establish a baseline, we computed the cosine similarity across prompts and obtained that the average similarity score hovers around 0.55 - 0.7, see Table \ref{inbetween_metrics}. This is again clearly a result of the custom nature of the LAMMPS input language, which is not fully captured by the cosine similarity metric. It is interesting to note that we find increasing variability in the scripts generated with increasing complexity of the task. This reflect the fact that more ways exist to set up a correct simulation but also that GPT-4 is less certain about the answer. This result suggests that the consistency score could be used by researchers as a surrogate for the accuracy of the script generated. We include pair-wise similarity comparisons as Figure S\mbox{\ref{figs3:cosine}}, and an average excluding self-similarity is added in Table \mbox{\ref{task1_metrics}}. Regarding their usefulness, it is clear that only a minor number of adjustments are needed to make these systems run.

\begin{table}[ht]
\caption{Cosine similarities for GPT-4 generated input files between different prompts describing a molecular dynamics simulation.}
\label{inbetween_metrics}
\vskip 0.3in
\begin{center}
\begin{tabular}{lcccr}
\toprule
 & Prompt 1 & Prompt 2 & Prompt 3 \\
\midrule
Prompt 1          & 0.87 &     &     \\
Prompt 2          & 0.72 & 0.85 &     \\
Prompt 3          & 0.56 & 0.55 & 0.81 \\

\bottomrule
\end{tabular}
\end{center}
\vskip -0.1in
\end{table}

% PROBLEMS IN PROMPTS 1/2
{\bf Common problems.} Inaccuracies in the input files for Prompts 1 and 2 are primarily due to a redundant \textit{replicate} command, which results in system sizes significantly larger than requested, see Figure \ref{fig1:annotatedcode}. As observed in Figure \ref{fig1:annotatedcode}, superfluous keywords were also often included in the command to initialize velocities, \textit{velocity create}. The \textit{mom/rot} keywords are used to zero out the linear and angular momentum of the newly created velocities. A uniform distribution was chosen to generate atomistic velocities in one of the files. While this would have a negligible effect on the results, a Maxwell-Boltzmann distribution is preferred. 
% PROBLEMS IN PROMPT 3
Prompt 3 introduces significant complexity. Although no input file generated by GPT-4 could be executed fully without adjustments, an input file matching the prompt can be obtained with minor modifications, merging the best parts of several files. This task requires creating a complex system including a target and a projectile and performing two distinct processes, a thermalization step followed by an impact simulation. GPT-4 consistently switches from an isothermal simulation (NVT ensemble) for equilibration to an adiabatic one (NVE) for the spall portion. 9 out the 10 input files generated by GPT-4 create separate projectile and target systems but none of the them are separated by the requested 1.5 nm. Setting the dimensions of the simulation box and boundary conditions requires an understanding of the impact simulations. Following the initial compression, the system expands. For impact velocities above a threshold value spall failure occurs in tension. Thus, the system size along the impact direction should enable this process or a spacial-purpose, adjustable, boundary condition should be selected. This is not accomplished in any of the input files but is easily fixed. 
An important step in the simulation workflow is the addition of a relative impact velocity of 2 km/s z-axis on top of thermal velocities of atoms in the projectile portion of the system. This requires thermalization of the system, which is generally achieved by these files (in 8 out of 10 cases), but only 2 out of the remaining 8 input files correctly add the desired impact velocity on top of the thermal ones (this requires the addition of \textit{sum yes} in the \textit{velocity} command). The remaining 8 input files replace the thermal velocities with the collective velocity. 

{\bf In-line comments}. In addition to the generation of correct or nearly correct files, useful warnings, and in-line comments coupled with the model's ability to follow up on additional questions to clarify or expand upon the generated code represent a powerful tool that can speed up the on-boarding of new users and accelerate the work of experts. Moreover, such explanations of the code's logic can aid in enhancing the user's comprehension of the LAMMPS input file and its implementation. 

\subsection*{Task 2: Reproducibility of published results}
\label{task2_section}

Task 1 demonstrated the GPT-4 model's ability to generate useful LAMMPS input files. In Task 2, we examine its potential to enhance the reproducibility of published computational studies. To this end we used GPT-4 to generate input files based on the simulation details provided in the Methods section, or similar, of publications and compare the GPT-4 files with those either originally shared with the publication (in the first example) or one created for this publication (second example). 

\subsubsection*{Montes de Oca Zapiain et al (2022). npj Computational Materials, 8(1), 189. \cite{montes2022training}}

In their study, Montes de Oca Zapiain and collaborators explored various methods for generating diverse training sets to improve the transferability of machine learning interatomic potentials. A key part of their analysis involves executing a LAMMPS simulation to assess energy conservation of an MD run. We took some of the details from this publication's methods to evaluate the GPT-4's (Version March 23, 2023) capabilities to reconstruct the simulation.  We extracted the following excerpt from the Methods section of their publication, which outlines the parameters for the MD simulations: 

\begin{promptquotebox}{Prompt: Montes de Oca Zapiain et al (2022).}
\textit{All the NVE simulations were initialized with 16 tungsten atoms in a BCC lattice with periodic boundary conditions, equilibrated at a temperature of 3000 K, close to the melting point. Each simulation was run for a total simulation time of 7.5 ps and the number of time samples n was 1000. All calculations were performed using the publicly released version of LAMMPS from November 2021. In addition to the base code, LAMMPS was compiled with the ML\-IAP, PYTHON, and ML\-SNAP packages. The MLIAP\_ENABLE\_PYTHON and BUILD\_SHARED\_LIBS compile flags were set.}
\end{promptquotebox}

Figure \ref{fig3:mitch_paper} compares the input file provided as supplemental information to the article and the input file generated using GPT-4. The name of the potential from the publication, was replaced with a placeholder for clarity. 

\begin{listing}[H]
\begin{minipage}[]{\linewidth}
\begin{multicols}{2}

\begin{minted}
[
frame=lines,
framesep=2mm,
baselinestretch=1.2,
bgcolor=LightGray,
escapeinside=||,
fontsize=\footnotesize,
linenos,
breaklines
]
{python}
variable dtfac index 1
variable dt equal 0.5e-3/|\$|{dtfac}
variable nthermo equal 15000*|\$|{dtfac}
variable nsteps equal 15000*|\$|{dtfac}
variable t index 3000
# sampling
variable nrepeat equal 250
variable nfreq equal round(|\$|{nthermo}/|\$|{nrepeat})
variable nrep equal 2
variable a equal 3.16
units metal
variable nx equal |\$|{nrep}
variable ny equal |\$|{nrep}
variable nz equal |\$|{nrep}
boundary p p p
lattice bcc |\$|a
region box block 0 |\$|{nx} 0 |\$|{ny} 0 |\$|{nz}
create_box 1 box
create_atoms 1 box
mass 1 183.84
velocity all create |\$|t 4928459 loop geom

timestep |\$|{dt}
# potential
pair_style mliap model mliappy <potential file> descriptor sna W.mliap.descriptor
pair_coeff * * W
run 0
# mean deviation
variable etot0 equal etotal
variable etotdelta equal abs(etotal-|\$|{etot0})/atoms
fix avede all ave/time |\$|{nfreq} |\$|{nrepeat} |\$|{nthermo} v_etotdelta
thermo_style custom step temp epair ke etotal press v_etotdelta f_avede
thermo |\$|{nthermo}
# run
fix mynve all nve
run |\$|{nsteps}
\end{minted}

\begin{minted}
[
frame=lines,
framesep=2mm,
baselinestretch=1.2,
bgcolor=LightGray,
escapeinside=||,
fontsize=\footnotesize,
linenos,
breaklines
]
{python}
||










units metal
atom_style atomic


boundary p p p
lattice bcc 3.1652
region box block 0 2 0 2 0 2
create_box 1 box
create_atoms 1 box

velocity all create 3000 87287 rot yes dist gaussian
timestep 0.001

pair_style mliap model |\st{nn} \textcolor{red}{mliappy}| model.mliap descriptor |\textcolor{red}{sna}| mliap.sna
pair_coeff * * W









thermo 1000

fix 1 all nve
run 7500

\end{minted}
\end{multicols}
\end{minipage}

\caption{Comparison between (a) published LAMMPS input file from Montes de Oca Zapiain et al \cite{montes2022training} and (b) generated LAMMPS input file (\#1) using GPT-4 from their methods description.}
\label{fig3:mitch_paper}
\end{listing}

Similar to the results in Task 1, the input file generated effectively captures the crystal structure, boundary conditions, lattice parameter, and initial material system. The command to generate velocities includes the \textit{rot} and \textit{dist} keywords. A significant difference between the two input files is the definition of variables in the authors' version, which is done to simplify the process of running multiple simulations. This was not included in the prompt, and, in fact, a simple substitution with the values declared results in strikingly similar files. GPT-4 was able to generate a \textit{pair\_style} command for a machine-learned interatomic potential (MLIAP); however, it included a couple of mistakes as marked in Figure \mbox{\ref{fig3:mitch_paper}}. Interestingly, GPT-4 made a reference to descriptors in a file type .sna, potentially making use of the compilation packages included in the prompt to identify the descriptors as SNAP. However, it missed a flag in the \textit{descriptor} keyword to specify them as such. The extent to which GPT-4 can capture the nuances related to the descriptors for the MLIAP is unclear. Commands for the calculation of an analysis conducted in the original publication (mean deviation of total energy per atom), which was included in the authors' input file, was absent in both the prompt and, therefore, missing in the GPT-4 input file. A small discrepancy between the GPT-4 and the authors' file relates to the simulation timestep. The original publication has 15,000 steps of 0.0005 ps, and GPT-4 has 7,500 steps of 0.001 ps. Nonetheless, both of them are correct and result in 7.5 ps of simulation. In summary, while none of the input files run without modification, a simple change in the definition of the interatomic potential, marked in red in Figure \ref{fig3:mitch_paper}, results in the desired simulation.

\subsubsection*{Strachan and Dorso (1998). Physical Review C. 58(2), R632. \cite{strachan1998caloric}}

In this publication, Strachan and Dorso explored the phenomena of multifragmentation of small many-body systems. This paper was selected for two reasons, it was published before the development of LAMMPS \cite{plimpton1995fast} and it involves a 2D simulation, a seldom used capability of LAMMPS. We extracted some of the details from this publication's methods to evaluate the GPT-4 model's capabilities for reproducibility. GPT-4 model for this prompt in Task 2 was Version May 24, 2023. We compiled the following prompt that describes the parameters needed for the molecular dynamics simulations: 

\begin{promptquotebox}{Prompt: Strachan and Dorso (1998).}
\textit{We simulate the atomistic two-dimensional system formed by N = 100 Lennard Jones particles interacting via the two body potential: $V(r) = 4 *\epsilon *[(\sigma/r)^{12} - (\sigma/r)^{6} - (\sigma/(r_c))^{12} + (\sigma/(r_c))^{6}]$ where $r_{c}$ is the cutoff radius, the potential is taken as zero for $r \geq r_{c}$. We took $r_{c} = 3*\sigma$. The unit of time and energy are $t_0 = \sqrt{\sigma^2*m/(48*\epsilon)}$  and $\epsilon$, respectively. The time step that we used for the integration of the equation of motion was fixed to $t_{int} = 0.0025*t_0$, in this way the energy and total angular momentum is conserved better than 0.01\%. For the study of multifragmentation we analyzed the time evolution of compressed and hot drops. The initial configurations are obtained by cutting a circular drop from a thermalized two-dimensional Lennard-Jones periodic system. These initial configurations are macroscopically characterized by the energy E, the density $\rho$ (of the periodic system), and the number of particles which is fixed to N=100. We divide the system in ten concentric circular regions, centered in the c.m. of the system, and we calculate the mean radial velocity of each circular ring. The temperature of each circular region is defined as the velocity fluctuations around the measured collective motion (mean radial velocity) of the corresponding ring. The total kinetic energy is the sum of the one corresponding to the expansion and that associated to the local temperature. Run simulations for the following sets of parameters $(\rho=0.75*(1/\sigma^2)$ and $E=-1.1*\epsilon)$; $(\rho=0.75*(1/\sigma^2)$ and $E=-0.55*\epsilon)$; $(\rho=0.75*(1/\sigma^2)$ and $E=0.8*\epsilon)$; $(\rho=0.8*(1/\sigma^2)$ and $E=-0.3*\epsilon)$.}
\end{promptquotebox}

Figure S\ref{figs2:strachandorso} includes the LAMMPS input file generated from the text description from the publication by GPT-4. GPT-4's input files correctly set up a 2D system and select a single layer of a hexagonal lattice as the initial conditions. This is a closed packed arrangement in 2D and a good choice to start the simulations. All files include {\it for} loops to explore the various temperatures and densities requested and the majority set the correct density. While handling of loop variables often need correction, the input files provided an excellent starting point. This task requires two simulations, the equilibration of a high-temperature liquid sample followed by cutting a circular droplet and following their expansion with adiabatic MD. The second step is not included in any of the input files but is a relatively simple addition. 

\subsection*{Task 3: Human description for publications from LAMMPS input files}
\label{task3_section}

Tasks 1 and 2 focused on GPT-4's ability to generate useful LAMMPS input files. Now, we aim to assess its ability generate a human-interpretable description of an MD simulation from a LAMMPS input file. The goal is to evaluate its potential to improve the quality and accuracy of research reports and publications via complete and accurate descriptions. The prompt used in our first attempt, see Figure \ref{methods_detailed_GPT-4},  specified our intent to use the description in the methods section of a publication and provided the LAMMPS input file provided in Ref. \cite{montes2022training} and used in Task 2. As shown in Figure \ref{methods_detailed_GPT-4}, GPT-4 provides a detailed description of the task including implementation details such as variable names. The description goes significantly deeper that the comments in the file and indicates that GPT-4 understands the sequence of steps used in setting up an MD simulations. Note that it even described the calculation of the deviation of total energy. We believe that such a complete and detailed explanation would benefit a researcher unfamiliar with LAMMPS in understanding the simulation. Such a description is overly detailed to be included in the methods section of an academic publication. 

In a second test, we adjusted the query statement to include: 1) a target text length; 2) context pertaining to the expectations of such descriptors in academic journals; and 3) a directive to refrain from referencing LAMMPS commands directly.  We used GPT-4 version May 3, 2023 for the detailed reconstruction, and version May 12, 2023 for the short reconstruction. As shown in Figure \ref{methods_general_GPT-4}, GPT-4 demonstrates its capacity to write a simulation description that, in the author's opinion, would be suitable for an academic publication with minor edits. The main shortcoming is the mislabeling of the potential used for the MD simulation. MLIAP is described as ``smooth atomic neighbor environment representation" instead of the correct spectral neighbor analysis potential (SNAP). In regards to conventions and clarity, the description lacks a simulation time, it instead lists the timestep and number of steps. Additionally, it is vague in phrases such as ``[properties]... were computed and recorded at regular intervals" and ``results averaged over specificed time invervals". Overall in this attempt, GPT-4 is unable to fully filter out redundant information (i.e. explanations of the ensemble), it brings us much closer to a fully automated description with only the LAMMPS input file as an input. Providing the model with the entire manuscript might potentially result in a richer description. It's important to note that GPT-4 still lacks a full understanding of machine-learned potentials and descriptors, as it mistakenly classified them as Smooth Overlap of Atomic Positions (SOAP) descriptors rather than the correct spectral neighbor analysis potential (SNAP) descriptors. This inaccuracy is present in both descriptors and is consistent with the mistakes seen on Task 2. 

\begin{table}[ht]
\caption{Performance metrics for GPT-4 generated descriptions from the LAMMPS input script provided in Montes de Oca Zapiain et al (2022) \cite{montes2022training}.}
\label{task3_metrics}
\vskip 0.3in
\begin{center}
\begin{tabular}{lccr}
\toprule
Metric & Short Prompt & Detailed Prompt \\
\midrule
Script Validity    & 0/10 & 0/10 \\
Simulation Accuracy & 0/10 & 0/10\\
Consistency & 0.79 & 0.84 \\
Usefulness &  0.63 & 0.85 \\

\bottomrule
\end{tabular}

\end{center}
\vskip -0.1in
\end{table}

\begin{listing}[H]

\begin{promptbox}
\textit{\textbf{ 1) Write a detailed methods section for an academic publication in computational materials science based on the following LAMMPS script: Script }}
\begin{GPT-4}

In this study, a molecular dynamics simulation was performed using the Large-scale Atomic/Molecular Massively Parallel Simulator (LAMMPS) to investigate the mechanical properties of a computational material. The following methodology was applied:

1. Simulation parameters: The timestep (variable dt) was set to 0.5e-3/1 (1 being the dtfac index), and the total number of timesteps (variable nsteps) was set to 15,000. The temperature (variable t) was maintained at 300 K throughout the simulation. The thermodynamic data was output every 15,000 timesteps (variable nthermo). The simulation box was subjected to periodic boundary conditions in all three dimensions (x, y, and z).

2. Sampling parameters: The simulation was repeated 250 times (variable nrepeat), with the average thermodynamic properties calculated every 60 timesteps (variable nfreq, derived from nthermo/nrepeat). The system size was set by replicating the unit cell twice in each dimension (variable nrep), creating a total of eight unit cells.

3. Lattice structure and box creation: A body-centered cubic (bcc) lattice with a lattice constant of 3.16 Å (variable a) was used. The simulation box (region box) was created based on the dimensions of the replicated unit cells (nx, ny, and nz), which were derived from the nrep variable. A single atom type was assigned to the simulation box with a mass of 183.84 atomic mass units.

4. Atom velocities: The initial velocities of all atoms were generated using a Gaussian distribution corresponding to the specified temperature (300 K) and a random seed number (4928459).

5. Interatomic potential: The Machine Learning-Interatomic Potential (MLIAP) was employed in the study using a model (model name) and a descriptor based on the Smooth Overlap of Atomic Positions (SOAP) representation. The pair\_style and pair\_coeff commands were used to define the interatomic potential and coefficients, respectively.

6. Energy deviation calculation: Before running the simulation, the initial total energy of the system (etot0) was calculated. During the simulation, the mean energy deviation (etotdelta) was computed by taking the absolute difference between the current total energy and the initial total energy, and dividing it by the number of atoms.

7. Thermodynamic output: The simulation output included the step number, temperature, potential energy (epair), kinetic energy (ke), total energy (etotal), pressure, mean energy deviation (etotdelta), and average energy deviation (f\_avede) over the specified nfreq.

8. Integration: The simulation was integrated using the NVE (microcanonical) ensemble, in which the number of particles (N), volume (V), and total energy (E) are conserved. The simulation was run for the specified number of timesteps (nsteps) after setting up all the parameters.

\end{GPT-4}
\end{promptbox}\par\bigskip
\caption{``Detailed" description generated from the LAMMPS input script provided in Montes de Oca Zapiain et al (2022) \cite{montes2022training}.}
\label{methods_detailed_GPT-4}
\end{listing}

\begin{listing}
\begin{promptbox}
\textit{\textbf{ 2) Write a very short methods section for an academic publication in computational materials science based on the following LAMMPS script.  Most academic journals only require the minimum amount of details required for reproducibility. The text should avoid referencing keywords specific to LAMMPS and discuss the simulation in more general terms. Script }}
\begin{GPT-4}
In this computational materials science study, molecular dynamics simulations were performed to investigate the properties of a tungsten system. The system was modeled using a body-centered cubic (bcc) lattice with a lattice constant of 3.16 Å. A cubic simulation box with periodic boundary conditions in all three dimensions was created, and the size of the simulation box was determined by replicating the unit cell in the x, y, and z directions. The tungsten atoms were assigned initial velocities based on a Boltzmann distribution at 300 K.

A machine-learned interatomic potential (MLIAP) model was employed to describe the interactions between the tungsten atoms. The MLIAP model was trained using a descriptor based on the smooth atomic neighbor environment representation. The time step for the simulations was set to $0.5x10^{(-3)}$ ps, and the total number of steps was 15,000. The simulations were run under the NVE ensemble, where the number of particles, volume, and energy were kept constant.

Thermodynamic properties, including temperature, potential energy, kinetic energy, total energy, and pressure, were computed and recorded at regular intervals. The mean deviation in total energy per atom was calculated during the simulations, with results averaged over specified time intervals to obtain time-averaged quantities.
\end{GPT-4}
\end{promptbox}\par\bigskip
\caption{``Short" description generated from the LAMMPS input script provided in Montes de Oca Zapiain et al (2022) \cite{montes2022training}.}
\label{methods_general_GPT-4}
\end{listing}

In this task, we evaluate the model's ability to retrieve the original LAMMPS input file from the description itself provided. We generate 10 independent files from the text given by GPT-4. Results shown in Table \ref{task3_metrics} demonstrate that the power of descriptive explanations when using LLMs. A breakdown of the pairwise comparisons with the original input file is included as Figure S\ref{figs3:cosine}.

\section*{Conclusions}
\label{conclusions_section}

This paper studied GPT-4's ability to create LAMMPS input files from technical descriptions of computational tasks and the inverse problem of explaining, in technical English, computational tasks from LAMMPS input files. While human intervention is required except for the simplest tasks, we find GPT-4 to be a powerful tool to train beginners in MD simulations with LAMMPS, accelerating the development of LAMMPS input files. GPT-4's ability to act as interpreter between technical writing and code or files can accelerate progress in computational materials science and reduce the time required to reproduce results or create new tasks. We believe the our results indicate the need to reconsider education and training of researchers increasing the focus on the why and what of computational tasks and reducing the focus on how. While our results focused on MD simulations and LAMMPS, we believe our results can be extrapolated to other fields and codes with communities that actively share scripts and files online.  

\section*{Code and Data Availability}
\label{availability_section}

Input files generated by GPT-4 are available in: \\https://github.itap.purdue.edu/StrachanGroup/OpenAI\_LAMMPS. Files related to the original publications discussed in this work are available in their original reference.

\section*{Acknowledgements}

This effort was supported by the US National Science Foundation FAIROS program, award 2226418.

\section*{Ethics declarations}

\subsection*{Conflict of interest}

On behalf of all authors, the corresponding author states that there is no conflict of interest.

\subsection*{Author Information}

\textbf{Corresponding Author}

\textbf{Alejandro Strachan} - School of Materials Engineering and Birck Nanotechnology Center, Purdue University, West Lafayette, Indiana 47907; Email: strachan@purdue.edu \\

\bibliographystyle{unsrt}
\bibliography{references.bib}

\renewcommand\listingscaption{Figure S}
\setcounter{listing}{0}
\renewcommand\figurename{Figure S}
\setcounter{figure}{2}

\section*{Supplemental Information: GPT-4 as an interface between researchers and computational software: improving usability and reproducibility}

\begin{listing}[H]
\begin{minted}
[
frame=lines,
framesep=2mm,
baselinestretch=1.2,
bgcolor=LightGray,
escapeinside=||,
fontsize=\footnotesize,
linenos
]
{python}
|\colorbox{green}{units metal}|
|\colorbox{SpringGreen}{dimension 3}|
|\colorbox{green}{boundary p p p}|

|\colorbox{green}{atom\_style atomic}|
|\colorbox{green}{lattice fcc 3.52}|

|\colorbox{green}{region box 0 10 0 10 0 10} \colorbox{SpringGreen}{units lattice}|
|\colorbox{green}{create\_box 1 box}|
|\colorbox{green}{create\_atoms 1 box}|

|\colorbox{red}{replicate 10 10 10}|

|\colorbox{green}{pair\_style eam/fs}|
|\colorbox{green}{pair\_coeff * *} \colorbox{yellow}{<potential file name>} \colorbox{green}{Ni}|
|\colorbox{SpringGreen}{mass 1 58.69}|

|\colorbox{green}{velocity all create 600} \colorbox{yellow}{<random seed>} \colorbox{SpringGreen}{mom yes} \colorbox{yellow}{rot yes} \colorbox{yellow}{dist gaussian}|

|\colorbox{green}{fix 1 all npt temp 300.0 2500.0 0.1 iso} \colorbox{yellow}{1 1 1}|

|\colorbox{green}{thermo} \colorbox{yellow}{<thermo steps>}|
|\colorbox{green}{thermo\_style} \colorbox{yellow}{<custom step temp pe etotal press vol>}|

|\colorbox{green}{timestep 0.001}|
|\colorbox{red}{run 500000}|
|\colorbox{SpringGreen}{unfix 1}|

|\colorbox{SpringGreen}{dump, print, write\_data, ...}|

\end{minted}
\caption{Annotated example of a LAMMPS input script from Prompt 2. \colorbox{green}{Always added correct commands}. \colorbox{SpringGreen}{Often added correct commands that match default values}. \colorbox{yellow}{Values that were generated with different options in some scripts}. \colorbox{red}{Incorrect commands}.  }
\label{figs1:annotatedcodeP2}
\end{listing}

\begin{listing}[H]
\begin{minipage}[]{\linewidth}
\begin{multicols}{2}
\begin{minted}
[
frame=lines,
framesep=2mm,
baselinestretch=1.2,
bgcolor=LightGray,
escapeinside=||,
fontsize=\footnotesize,
linenos,
breaklines
]
{python}
# LAMMPS Input Script for Two-dimensional Lennard-Jones Particles

units lj
dimension 2
atom_style atomic
boundary p p f

variable sigma equal 1.0
variable epsilon equal 1.0
variable rc equal 3.0*|\$|{sigma}
variable N equal 100

lattice hex 0.75
region simbox block 0 25 0 25 -0.5 0.5
create_box 1 simbox
create_atoms 1 box

mass 1 1.0

pair_style lj/cut |\$|{rc}
pair_coeff 1 1 |\$|{epsilon} |\$|{sigma} |\$|{rc}

compute peratom all pe/atom
compute total all reduce sum c_peratom

neighbor 0.3 bin
neigh_modify delay 0 every 1 check yes

variable time_scale equal sqrt(|\$|{sigma}*|\$|{sigma}/(48.0*|\$|{epsilon}))
variable dt equal 0.0025*|\$|{time_scale}
timestep |\$|{dt}

# Simulations for different density and energy combinations
label loop_over_parameters
variable param_loop index 1 2 3 4
variable r loop 0.75 0.75 0.75 0.8
variable E loop -1.1 -0.55 0.8 -0.3

fix 1 all nve
variable T equal 1.0 # Specify target temperature here
fix 2 all langevin |\$|{T} |\$|{T} 1.0 987654

\end{minted}

\begin{minted}[
frame=lines,
framesep=2mm,
baselinestretch=1.2,
bgcolor=LightGray,
escapeinside=||,
fontsize=\footnotesize,
breaklines
]
{python}

thermo 100
thermo_style custom step temp epair etotal press

run 5000

unfix 1
unfix 2

# Calculate properties for ten concentric circular regions
variable rmax equal sqrt(|\$|{N}/(pi*|\$|{r}))
variable dr equal |\$|{rmax}/10.0
variable r_outer equal |\$|{dr}

label loop_over_regions
variable region_index loop 1 2 3 4 5 6 7 8 9 10

region circle_|\$|{region_index} cylinder z 0.0 0.0 0.0 0.0 |\$|{r_outer} INF INF
group circle_|\$|{region_index} region circle_|\$|{region_index}

variable v_rad_|\$|{region_index} atom sqrt((vx-vcm(circle_|\$|{region_index},vx)) * (vx-vcm(circle_|\$|{region_index},vx)) + (vy-vcm(circle_|\$|{region_index},vy)) * (vy-vcm(circle_|\$|{region_index},vy)))
variable temp_|\$|{region_index} equal temp(circle_|\$|{region_index})

compute ke_|\$|{region_index} circle_|\$|{region_index} ke/atom
compute sum_ke_|\$|{region_index} all reduce sum c_ke_|\$|{region_index}
variable total_ke equal |\$|{total_ke} + sum(c_sum_ke_|\$|{region_index})

next region_index
jump SELF loop_over_regions

next param_loop
jump SELF loop_over_parameters

\end{minted}
\end{multicols}
\end{minipage}

\caption{LAMMPS script generated by GPT-4 for the simulation conducted by Strachan and Dorso \cite{strachan1998caloric}}
\label{figs2:strachandorso}
\end{listing}

\begin{figure}[ht]
\centering
\begin{subfigure}{.3\textwidth}
\centering
\subcaptionOverlay{\includegraphics[width=\textwidth]{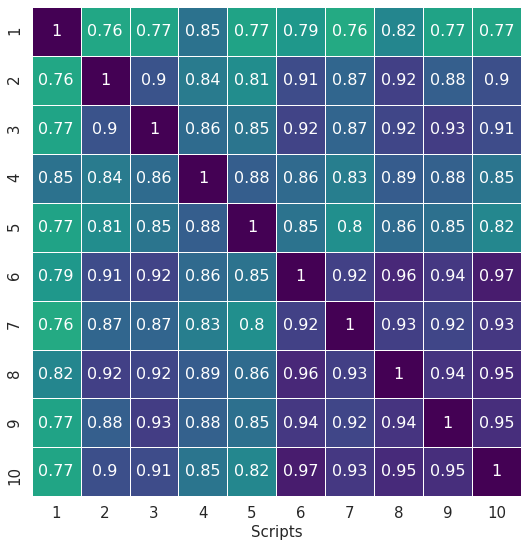}}
\end{subfigure}
\begin{subfigure}{.3\textwidth}
\centering
\subcaptionOverlay{\includegraphics[width=\textwidth]{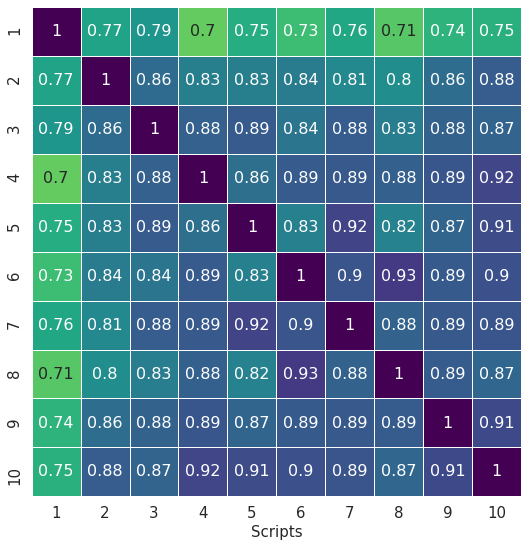}}
\end{subfigure}
\begin{subfigure}{.3\textwidth}
\centering
\subcaptionOverlay{\includegraphics[width=\textwidth]{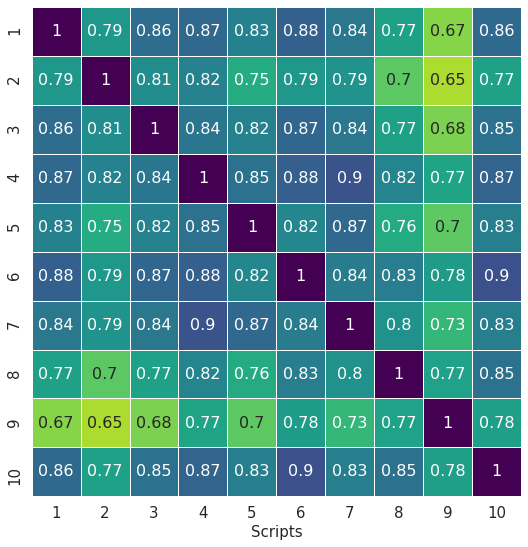}}
\end{subfigure}
\begin{subfigure}{.3\textwidth}
\centering
\subcaptionOverlay{\includegraphics[width=\textwidth]{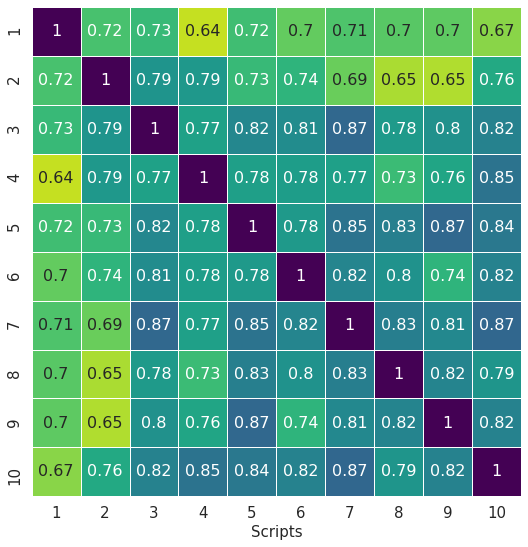}}
\end{subfigure}
\begin{subfigure}{.3\textwidth}
\centering
\subcaptionOverlay{\includegraphics[width=\textwidth]{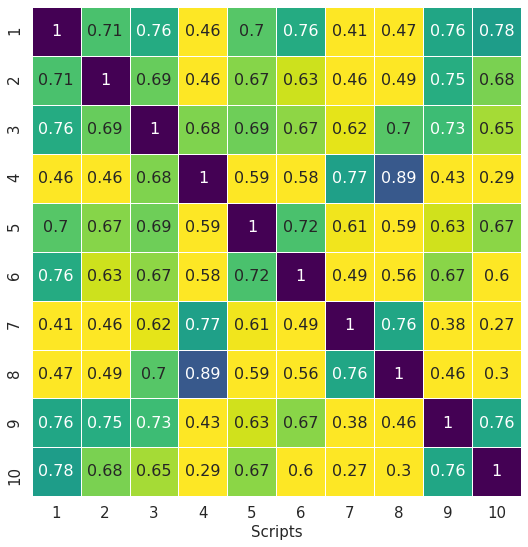}}
\end{subfigure}
\caption{Pairwise Cosine similarity for scripts generated in (a) Prompt 1 (b) Prompt 2 (c) Prompt 3, (d) Montes de Oca Zapiain et al\cite{montes2022training} and (e) Strachan and Dorso \cite{strachan1998caloric}. 
}
\label{figs3:cosine}
\end{figure}

\end{document}